\documentclass[a4paper,11pt]{article}
\pdfoutput=1
\usepackage{jheppub}
\usepackage{textcase}
\usepackage{graphicx,epsfig,psfrag,amssymb,hyperref}
\usepackage{multirow}
\usepackage{color,graphicx,epsfig,psfrag,amsmath,empheq}
\usepackage[dvipsnames]{xcolor}
\usepackage{bm}
\usepackage{mathrsfs,amsfonts,soul,color}
\usepackage[caption=false]{subfig}
\usepackage{enumitem}
\usepackage{placeins}
\usepackage{textcomp}
\usepackage{tcolorbox}
\usepackage{xspace}
\usepackage{float}
\usepackage[toc,page]{appendix}
\usepackage{wrapfig}
\usepackage[colorinlistoftodos]{todonotes}
\usepackage{siunitx}

\usepackage{mycommands}

\newcommand\snowmass{\begin{center}\rule[-0.2in]{\hsize}{0.01in}\\\rule{\hsize}{0.01in}\\
\vskip 0.1in Submitted to the  Proceedings of the US Community Study\\ 
on the Future of Particle Physics (Snowmass 2021)\\ 
\rule{\hsize}{0.01in}\\\rule[+0.2in]{\hsize}{0.01in} \end{center}}

\begin{document}


\title{The Mu3e Experiment} 

\author[1]{Gavin Hesketh,}
\author[2]{Sean Hughes,}
\author[3]{Ann-Kathrin Perrevoort,}
\author[2]{and Nikolaos Rompotis}
\author[]{on behalf of the Mu3e Collaboration\vspace{3mm}}
\affiliation[1]{University College London}
\affiliation[2]{University of Liverpool}
\affiliation[3]{Karlsruhe Institute of Technology}

\emailAdd{gavin.hesketh@ucl.ac.uk}

\abstract{

  The Mu3e experiment at the Paul Scherrer Institut will search for
  the lepton-number-violating decay \mueee, extending the sensitivity
  by four orders of magnitude compared to existing limits. This probe
  of new physics is complementary to the existing collider, dark
  matter and neutrino particle physics programmes, and part of a
  global programme investigating the charged lepton flavour sector. As
  well as the main \mueee search, Mu3e will also extend the
  sensitivity to low-mass dark photons, and additional
  flavour-violating decays involving long-lived or stable particles.

\snowmass

}
\maketitle

\clearpage

\section{Introduction}
\label{sec:intro}
Charged lepton flavour violating (CLFV) processes offer unique
discovery potential for physics beyond the Standard Model (BSM),
bringing sensitivity to new physics that is complementary to the
existing collider, dark matter and neutrino particle physics
programmes. The Mu3e experiment~\cite{Mu3e:2020gyw} at the Paul
Scherrer Institut (PSI) is part of a global programme of experiments
searching for the ``golden channels'' of CLFV in the muon
sector: \meg, \mueee, \muconv. This programme will bring a significant
increase in sensitivity compared to previous searches, probing new
physics mass scales up to $10^3$ - $10^4$~TeV.  Of the
experiments which will start taking data in the coming years, Mu3e is
the only one which will search for the decay \mueee
(MEG-II~\cite{MEG:2016leq,MEGII:2018kmf} targets the \meg signal,
while Mu2e~\cite{Mu2e:2014fns}, COMET~\cite{COMET:2018auw} and
DeeMe~\cite{Teshima:2019orf} will search for \muconv).

Flavour violation is a known feature of the quark and neutrino sectors
of the Standard Model. And while charged lepton flavour appears to be
conserved, it is not protected by any known global symmetry and occurs
at the 1-loop level via neutrino oscillation. In the Standard Model,
it is suppressed to an unobservably small rate
$\mathcal{O}(10^{-50}$)~\cite{petcov1977processes, Blackstone:2019njl} by factors of
$\left(\Delta m_{ij}^2/m_W^2\right)^2$ in those loops, where $\Delta
m_{ij}$ is the squared mass difference between the neutrino mass
eigenstates $i$ and $j$. However, significant enhancements to this
rate are predicted by many BSM scenarios, and the observation of any
CLFV signal would be the unambiguous observation of new physics.

The nature of any CLFV signal will depend on the underlying
physics. For example, in processes dominated by $\gamma$-penguin
diagrams, the \meg rate is expected to be higher than \mueee
or \muconv. However, the reverse is true if $Z$ or $H$-penguin
diagrams or tree level $Z^\prime$ or lepto-quark models
dominate. There is an extensive and growing literature exploring these
different scenarios, for example~\cite{Calibbi:2017uvl}, which can be
parameterised with a general effective operator approach, as used in
Section~\ref{sec:mueee}. However, the conclusion is clear: exploring all
three ``golden'' muon channels is essential.

The current best limit on \mueee was set by the SINDRUM
collaboration~\cite{SINDRUM:1987nra}, excluding a branching ratio over
\num{1.0e-12} at 90\% confidence level (CL).    
With a two-phase approach, Mu3e aims to extend the sensitivity to
$10^{-16}$. Phase-1, currently under construction, will utilise the
$\pi$E5 beam-line at PSI to study up to $10^8$ muon ($\mu^+$) stops
per second and reach a sensitivity of $2\times10^{-15}$ on the
branching fraction. An upgraded detector for Phase-2 plans to make use
of the High-Intensity Muon Beam upgrades at PSI to study $2\times10^9$
muon stops per second and reach the target sensitivity of $10^{-16}$
on the branching fraction.

Mu3e uses an innovative detector design, with
extremely low material budgets in order to minimise multiple
scattering for low energy electrons. It consists of four layers of
HV-MAPS silicon pixel sensors thinned to 50~$\mu$m, and scintillating
fibre and tile detectors providing sub-ns timing resolution. The
detector is surrounded by a solenoid magnet providing a 1~T field; electrons
from muon decays can pass through the detector several times as they
follow helical paths in this field. This significantly extends the
lever-arm for measurement, and hence the momentum resolution.

As well as the main \mueee search, Mu3e can search for signatures of
the form \mueX, where $X$ can decay to $e^+e^-$ promptly, after
travelling some distance, or escaping the detector before
decaying. This covers a range of ALP, familon, majoron and $Z^\prime$
models.

This White-paper will briefly cover the expected sensitivity of the various
searches possible at Mu3e, along with an overview of the
experimental design.


\section{The search for \mueee}
\label{sec:mueee}
To study the effects of different BSM scenarios on the decay
kinematics, and therefore detector acceptance, we will consider the
general effective Lagrangian proposed by Kuno and
Okada~\cite{Kuno:1999jp}:
\begin{align} \label{eq:lagrangian}
\mathcal{L}_{\mu \rightarrow eee} = -\frac{4G_F}{\sqrt{2}}\big[ & m_\mu A_R \; \overline{\mu_R} \sigma^{\mu \nu} e_L F_{\mu \nu}  +  m_\mu A_L \; \overline{\mu_L} \sigma^{\mu \nu} e_R F_{\mu \nu}  \notag \\
 & +  \;g_1 \; (\overline{\mu_R} e_L) \; (\overline{e_R} e_L) + g_2 \; (\overline{\mu_L} e_R)\; (\overline{e_L}e_R)  \notag   \\
 & + \;g_3 \;  (\overline{\mu_R} \gamma^\mu e_R) \; (\overline{e_R} \gamma_\mu e_R)  + g_4 \; (\overline{\mu_L} \gamma^\mu e_L) \; (\overline{e_L} \gamma_\mu e_L)   \notag  \\
 & + \; g_5 \;  (\overline{\mu_R} \gamma^\mu e_R) \; (\overline{e_L} \gamma_\mu e_L)  + g_6 \; (\overline{\mu_L} \gamma^\mu e_L) \; (\overline{e_R} \gamma_\mu e_R) \; + \; H.c.
\; \big] 
\end{align}
where the form factors $A_{R, L}$ describe tensor-type (dipole) couplings, which contribute to \mueee primarily through higher-order photon penguin diagrams; the other terms describe tree-level contact interactions with $g_{1,2}$ describing scalar-type and $g_{3-6}$ vector-type interactions.

\begin{figure}[ht]
    \centering
    \subfloat[]{\label{fig:limits1}
    \includegraphics[width=0.49\textwidth]{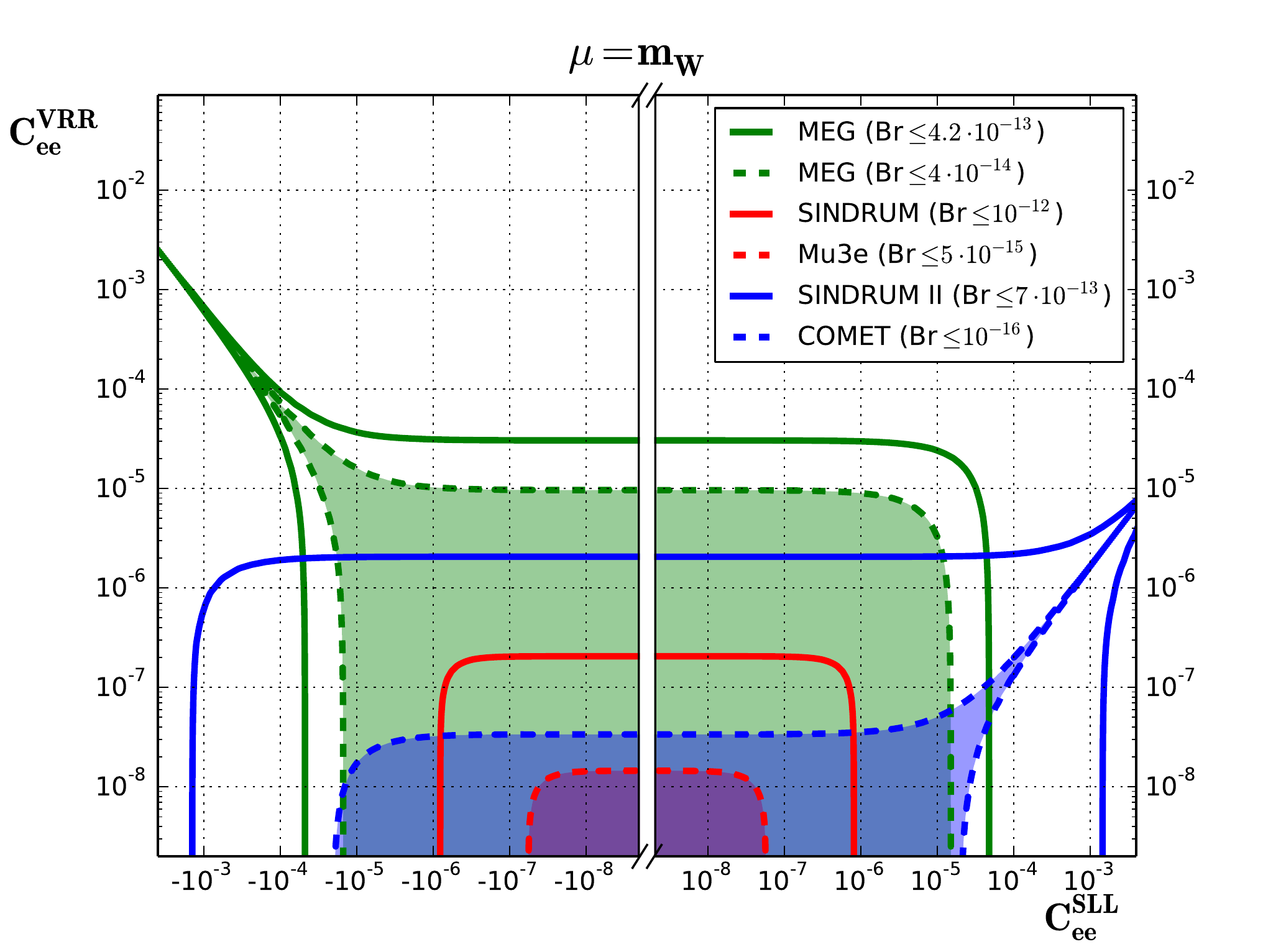}}
    \subfloat[]{\label{fig:limits2}
    \includegraphics[width=0.49\textwidth]{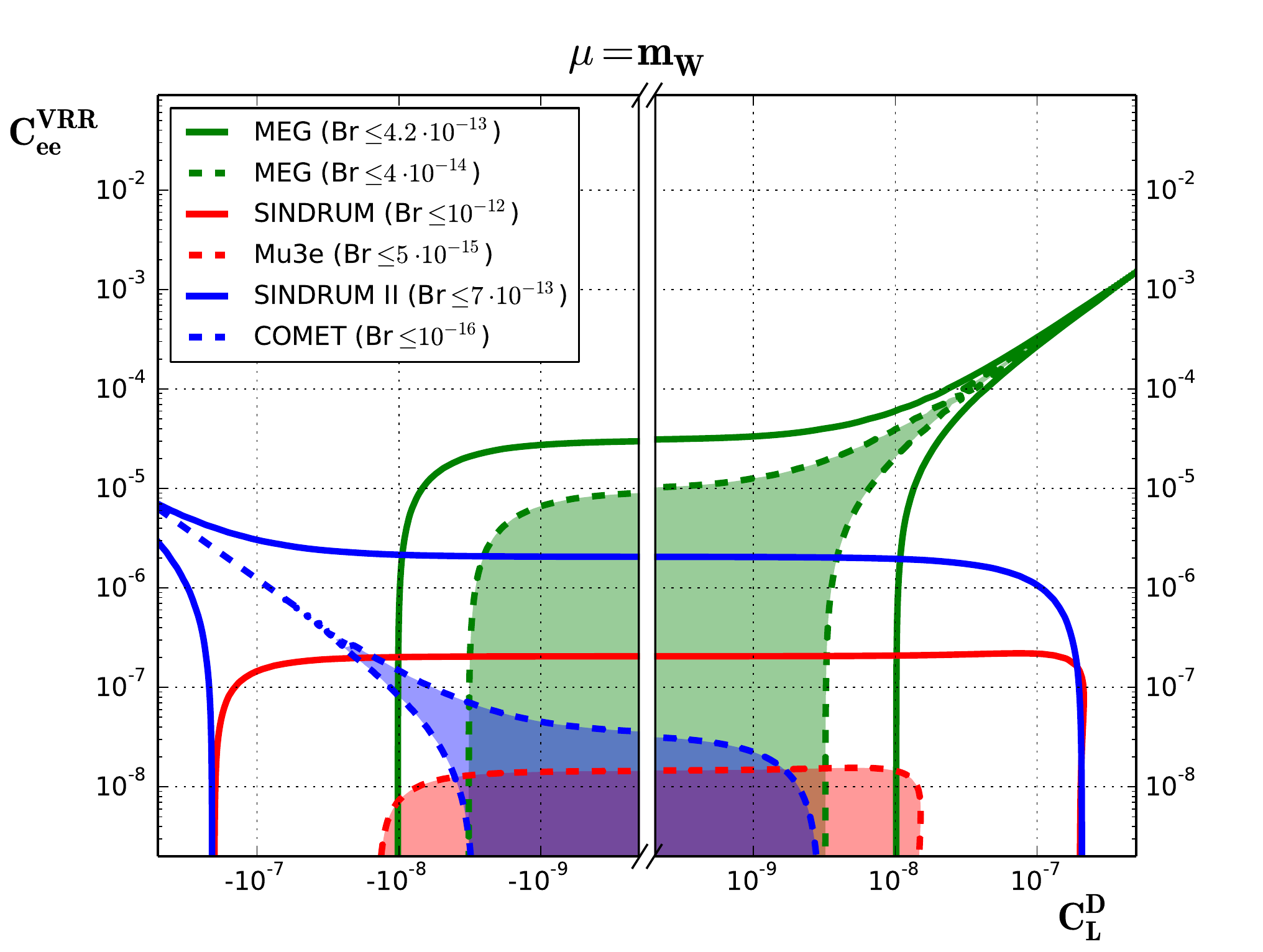}}
    \caption{Allowed regions (given at the scale $m_W$) in the \ref{fig:limits1} $C_{ee}^{VRR}$ - $C_{ee}^{SLL}$ plane, and the \ref{fig:limits2} $C_{ee}^{VRR}$ - $C_{L}^{D}$ plane. Existing (solid lines) and projected (dashed lines) are shown for \meg (green), \mueee (red) and \muconv (blue). Figures from~\cite{Crivellin:2017rmk}.} 
    \label{fig:limits}
\end{figure}

The expected constraints on these effective operators have been studied in detail in, for example,~\cite{Crivellin:2017rmk}. Figure~\ref{fig:limits} (from~\cite{Crivellin:2017rmk}) shows the allowed regions in two planes, defined by Wilson coefficients $C_{ee}^{VRR}$ (equivalent to $g_3$ in eq.~\ref{eq:lagrangian}), $C_{ee}^{SLL}$  (equivalent to $g_1$ in eq.~\ref{eq:lagrangian}), and $C_{L}^{D}$ (equivalent to $A_{R}$ in eq.~\ref{eq:lagrangian}).
It is worth noting that, for the purely leptonic contact-type interactions (Fig.~\ref{fig:limits1}), Mu3e is perhaps unsurprisingly the most sensitive. The other channels (\meg and \muconv) are also ``blind'' to certain regions of the parameter space due to cancellations, while this is never the case for \mueee. When considering the dipole-type interaction (Fig.~\ref{fig:limits2}), the different channels are more complementary. When considering the flavour dependnce in the contact-type interactions, the muon conversion experiments set more strungent limits than \mueee or \meg. The studies in~\cite{Crivellin:2017rmk} highlight the degeneracy between different terms in the effective Lagrangian in~\ref{eq:lagrangian}, and the need for inputs from all three muon decay channels in order to resolve such degeneracy in the event of any observation of a signal. 

\begin{figure}
    \centering
    \subfloat[]{\label{fig:mueeeAcceptance}
    \includegraphics[width=0.49\textwidth]{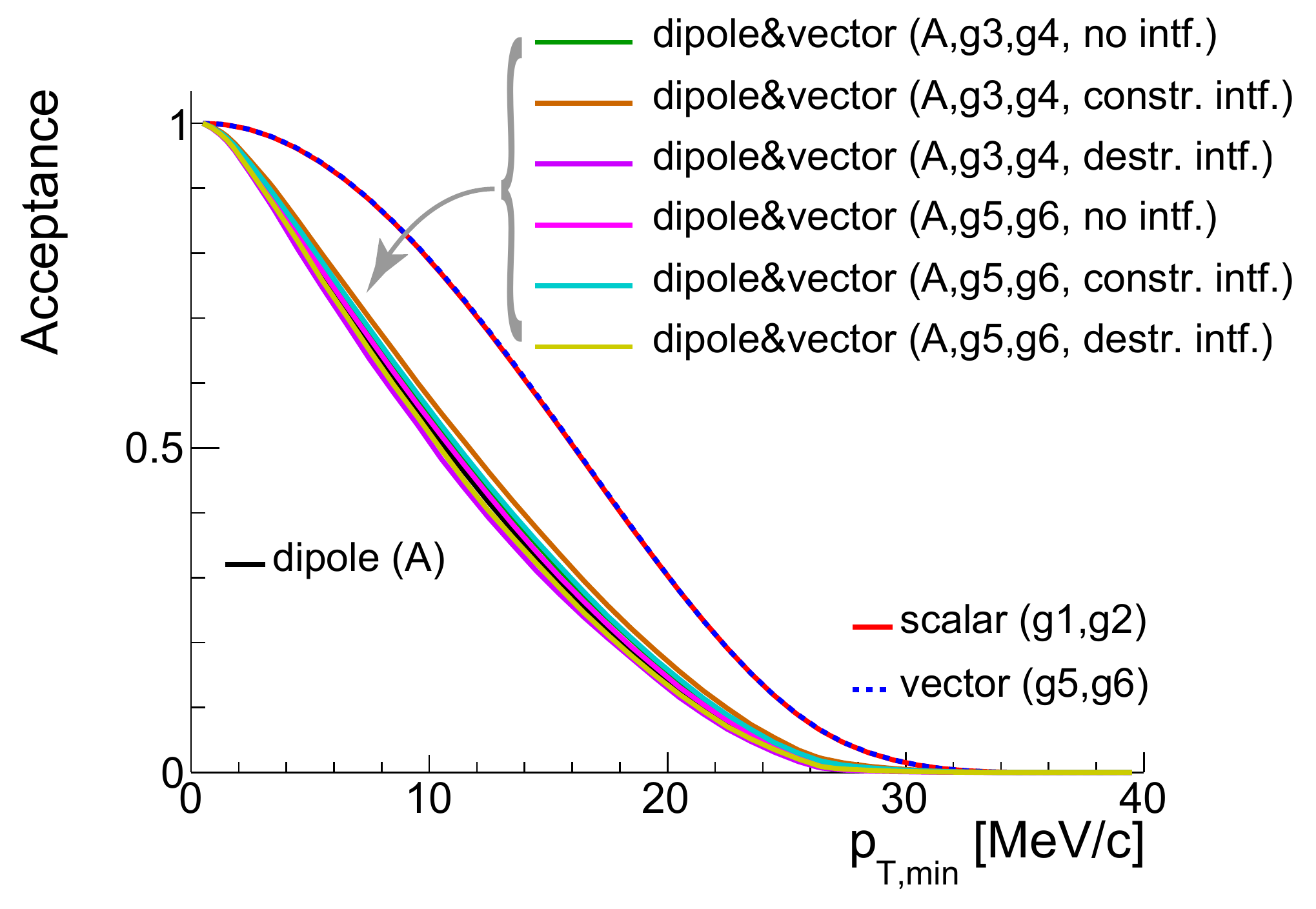}}
    \subfloat[]{\label{fig:mueeeEmax}
    \includegraphics[width=0.49\textwidth]{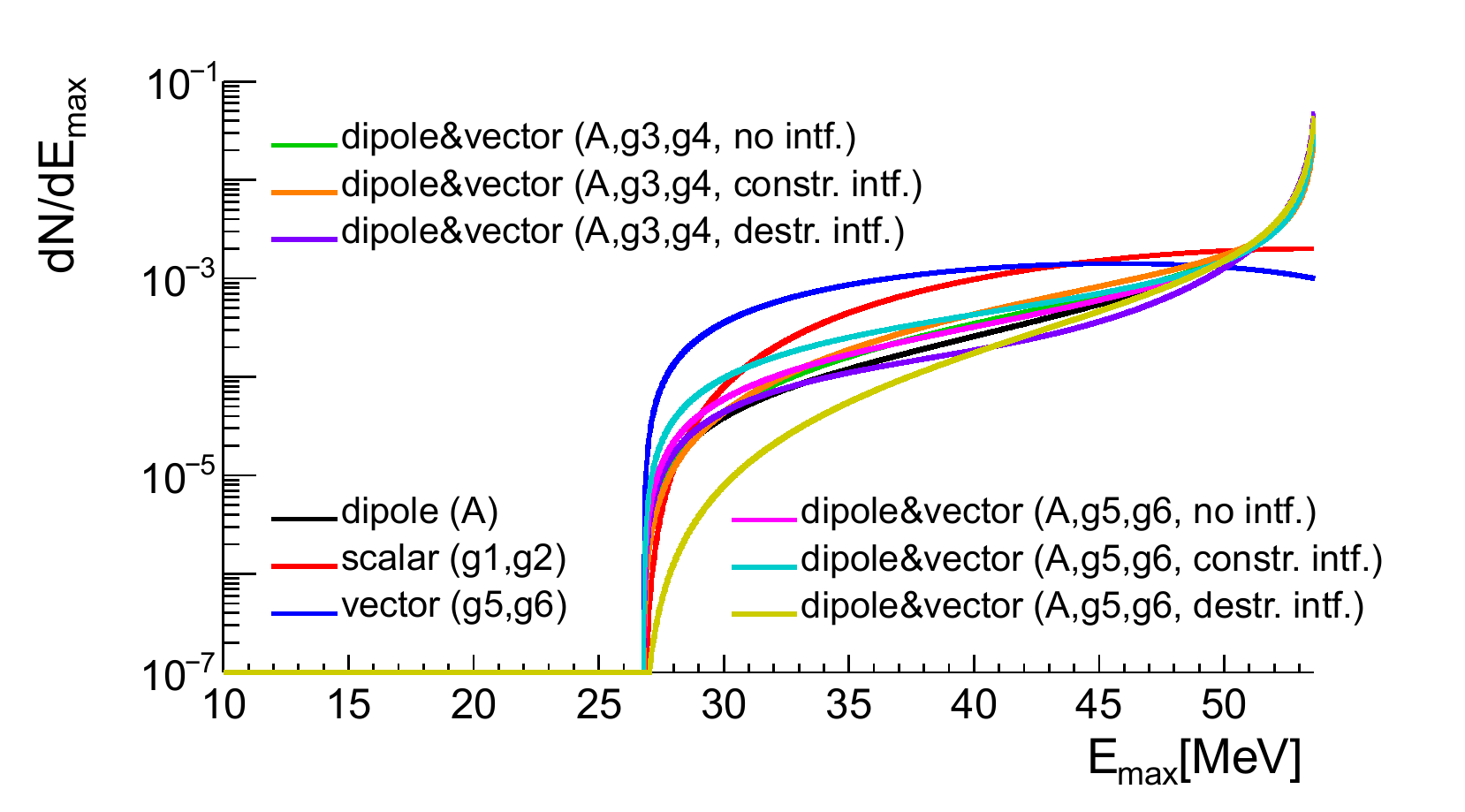}}
    \caption{\ref{fig:mueeeAcceptance} the acceptance, defined as the fraction of \mueee decays where all decay products have \pt greater than \ptmin. \ref{fig:mueeeEmax} the energy distribution of the highest energy decay product in \mueee decays. Both are shown for the range of effective operators defined by~\ref{eq:lagrangian}. Figures from~\cite{Mu3e:2020gyw}.} 
    \label{fig:mueeeKinem}
\end{figure}

The general effective Lagrangian in Eq.~\ref{eq:lagrangian} can also 
be used to study signal kinematics and provides an important input in
the design of the Mu3e experiment. Two important considerations are
the acceptance, defined as the fraction of \mueee decays
where all decay products have a transverse momentum \pt above some
minimal value \ptmin; and the energy distribution of the highest
energy decay product in \mueee decays. Both distributions are shown in
Fig.~\ref{fig:mueeeKinem} (taken from~\cite{Mu3e:2020gyw}) and
illustrate that maximising sensitivity requires the ability to
reconstruct electrons from half the muon mass down to as low as
possible in \pt.

The other key consideration in the \mueee search is backgrounds, which
fall into two categories: physics and combinatorics. The primary
physics background is \textit{internal conversion}:
\mueeenunu, which can be separated
from the signal by the presence of two neutrinos. Combinatorics arise
from the coincidence of one or more standard Michel decays with an
$e^-$ arising primarily from Bhabha scattering or radiative decay
($\mu^+\to\overline{\nu}_\mu e^+\nu_e\gamma$) with subsequent photon conversion
in the material of the detector. The three tracks in such cases
typically do not have a common origin, and are not coincident in time.

All backgrounds can be controlled using vertexing and kinematic
requirements, selecting three tracks consistent with $e^+e^-e^+$ from
a common origin, and with the reconstructed vertex momentum $<4$ MeV
and mass consistent with the muon mass (103 MeV$ < m_{eee} <$ 110
MeV). The combinatoric background can be further reduced by
coincident timing requirements, with time resolutions below 100~ps
required to obtain a two orders of magnitude reduction. These physics
requirements drive the design of the Mu3e experiment, which combines
an extremely low material budget tracking detector with scintillating
fibres and tiles for timing measurements. The Phase-1 Mu3e design is
briefly reviewed below, with full details in~\cite{Mu3e:2020gyw}.

In order to obtain optimal momentum and vertexing resolution on low
energy ($<m_\mu/2$) electrons, Mu3e consists of four layers of
High-Voltage Monolithic Active Pixel Sensors (HV-MAPS) surrounding the
muon stopping target. These sensors have a pixel size of
$80\times80$~$\mu$m$^2$, and are thinned to 50~$\mu$m giving a
thickness of $X/X_0 = 0.1$\% per layer when including the full
assembly. Between the second and third pixel layers, the scintillating
fibre detector provides the first time measurement. The fibre detector
consists of three layers of 250~$\mu$m diameter fibres, providing a
time resolution of 250~ps with an overall thickness of $X/X_0 =
0.2$\%.  The entire detector sits in a wide-bore (1~m diameter)
superconducting solenoid magnet providing a 1~T field.

\begin{figure}[ht]
    \centering
    \subfloat[]{\label{fig:schematicxy}
    \includegraphics[width=0.7\textwidth]{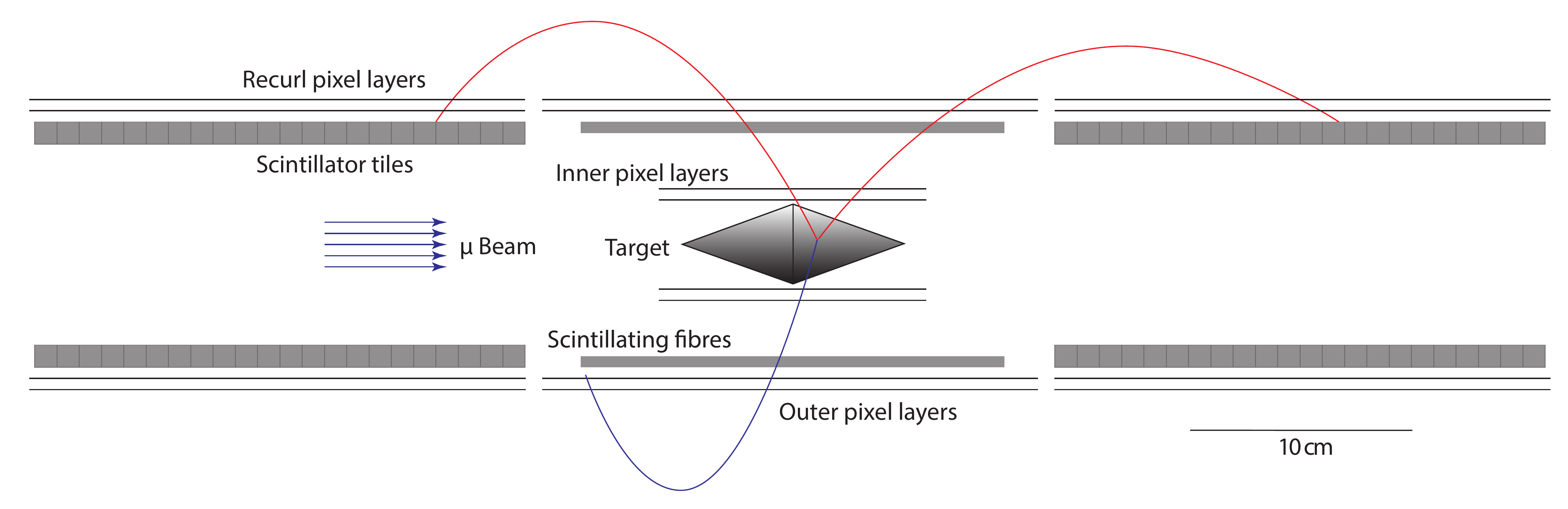}}
    \subfloat[]{\label{fig:schematicyz}
    \includegraphics[width=0.28\textwidth]{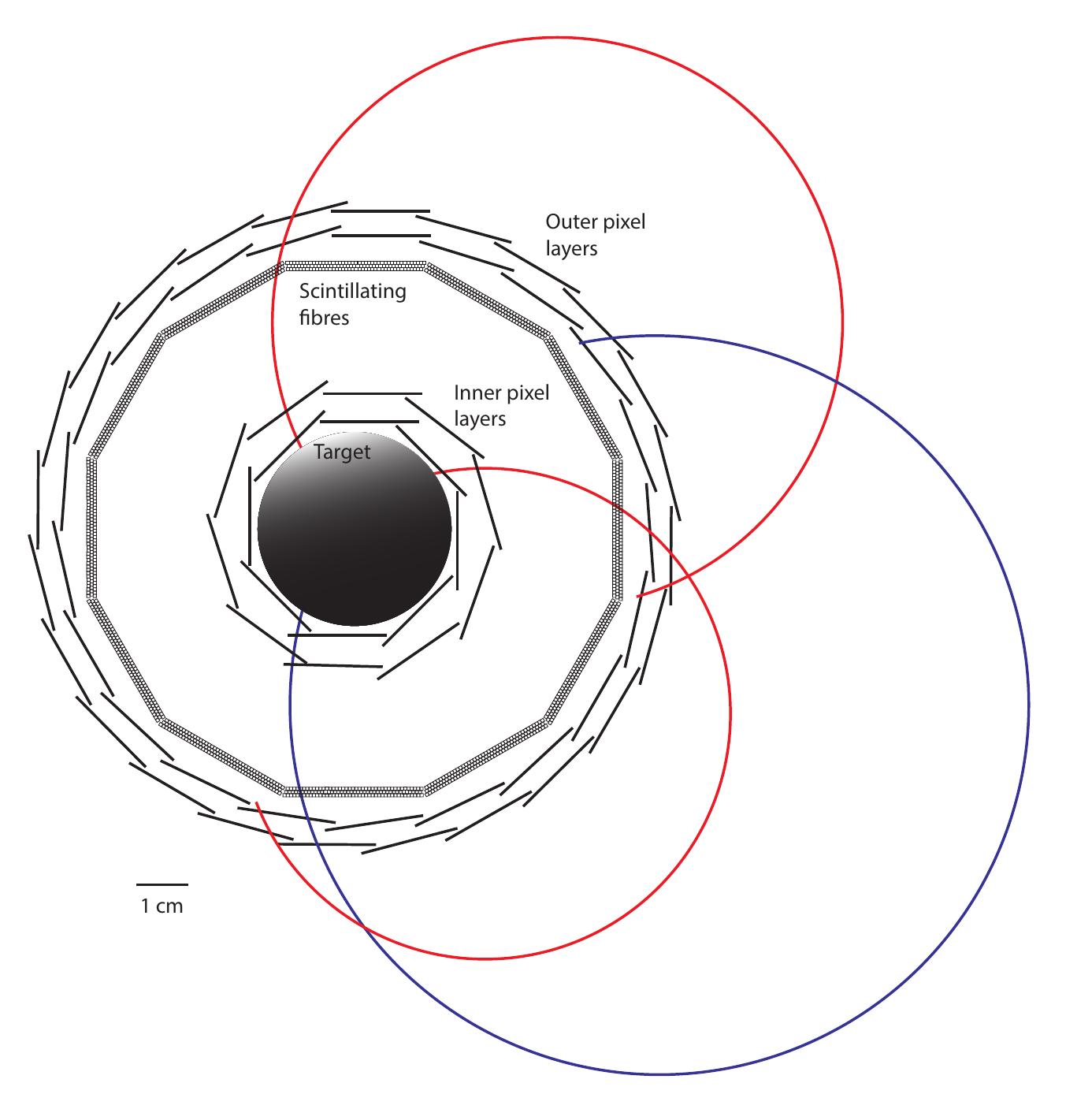}}
    \caption{The layout of the Phase-1 Mu3e detector. Figures from~\cite{Mu3e:2020gyw}.} 
    \label{fig:schematic}
\end{figure}

In this multiple-scattering dominated regime, optimal momentum
resolution is obtained for tracks which ``recurl'' in the magnetic
field and pass through the detector a second time. To increase the
acceptance for such tracks, upstream and downstream \textit{recurl
stations}, consisting of two further layers of pixel sensors
surrounding a layer of plastic scintillator tiles are
installed. Individual tiles measure $6.3 \times 6.2 \times
5\si{\mm^{3}}$, and are mounted in seven modules of 416 tiles each to
provide full azimuthal coverage in each recurl station. The size of the tiles 
is chosen to provide improved time resolution ($<50$~ps) compared to the fibre
detector, with a thickness which ensures high ($>99$\%)
efficiency. The additional scattering in the tiles is not relevant, as
only the pixel hits before reaching the tiles are used in track
fitting. Figure~\ref{fig:schematic} shows the layout of the Phase-1
Mu3e detector.

\begin{figure}[ht]
    \centering \subfloat[]{\label{fig:mass_plot} \includegraphics[width=0.49\textwidth]{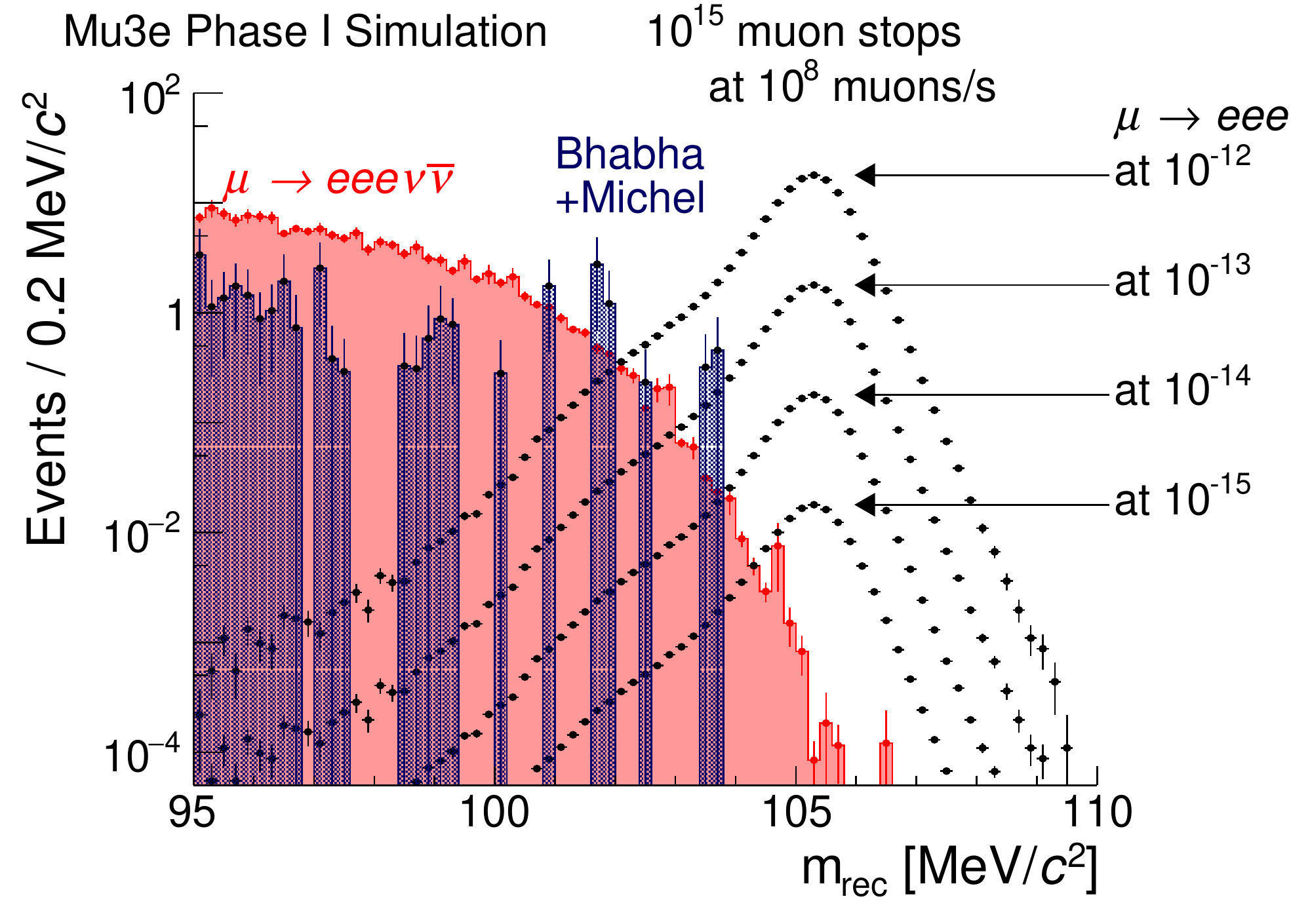}} \subfloat[]{\label{fig:BRsensitivity} \includegraphics[width=0.49\textwidth]{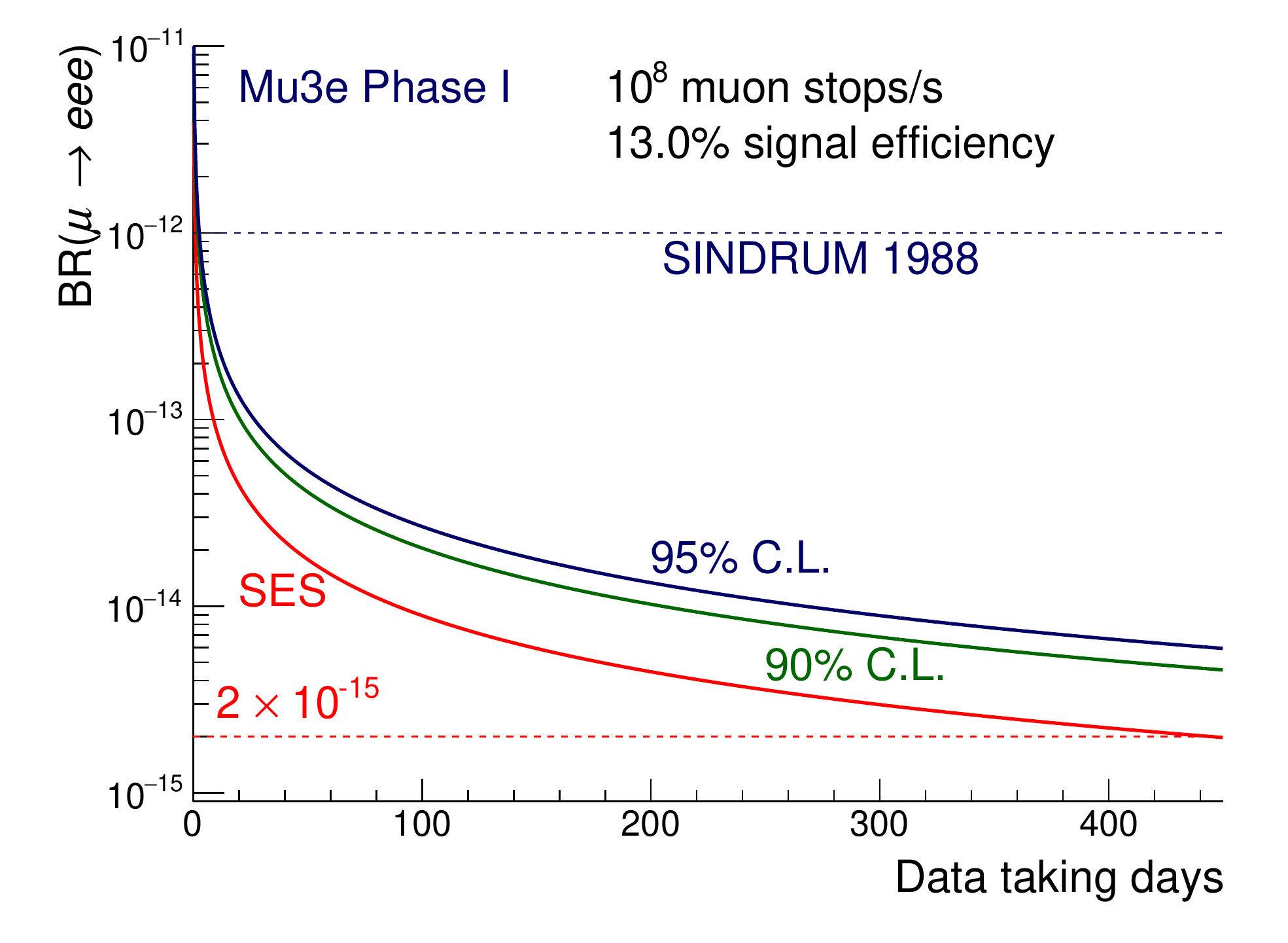}} \caption{\ref{fig:mass_plot}
    Simulation of the reconstructed vertex mass showing backgrounds
    and possible signal contributions. \ref{fig:BRsensitivity} the
    evolution of the Mu3e Phase 1 signal sensitivity with
    time. Figures from~\cite{Mu3e:2020gyw}}. \label{fig:sensitivity}
\end{figure}

The high rate of coincidences means that simply triggering on three
coincident tracks does not provide a sufficient reduction in data rate
to storage. Instead, Mu3e will operate with a continuous, triggerless
readout, with a fast online track reconstruction used to select events
containing three tracks consistent with a common vertex. The online
reconstruction is based on time-slices (``frames'') of the full
detector readout, with 4-hit tracks reconstructed in each frame on a
GPU farm. The full offline tracking and signal selection, based on
longer recurling (6 or 8 hit) tracks to optimise resolution, is then
carried out on the stored frames. Figure~\ref{fig:mass_plot} shows the
expected vertex mass distribution based on the full offline
selection. Figure~\ref{fig:BRsensitivity} shows the expected evolution
of sensitivity with running time; existing limits will be superseded
within days, and the target sensitivity reached with around 400 days
of data taking with a muon stopping rate of $1\times10^8$~s$^{-1}$.

\subsection*{Mu3e Phase-2}
To reach the final target sensitivity of $10^{-16}$ on the branching
fraction for \mueee, a higher rate of muon stops is required. The High
Intensity Muon Beam (HiMB) currently under study at PSI would deliver
a stopping rate of $2\times10^9$~s$^{-1}$, but is not expected to be
available before 2028. In order to deal with this higher stopping rate
and resultant higher occupancy and rate of coincident backgrounds,
upgrades to the Mu3e timing detectors are necessary, as well as
possible improvements to the pixel sensors to improve time resolution,
and extensions to the detector stations to increase the
acceptance. Such upgrades are currently under study.

\FloatBarrier

\section{The search for \mueX}
\label{sec:mueX}
In addition to searches for \mueee decays, Mu3e can also investigate
lepton-flavor-violating decays of the type \mueX, where $X$ denotes a
neutral light particle that escapes the experiment undetected. An
example for such a particle is the familon which arises as a
pseudo-Goldstone boson from an additional broken flavour
symmetry~\cite{Wilczek:1982rv}. The current strongest limits on the
branching ratio of \mueX are set by the experiment by Jodidio et al.\
at TRIUMF~\cite{Jodidio:1986mz} for massless $X$ with
$\BR(\mueX)<\num{2.6e-6}$ at 90\%\,CL, as well as the TWIST
experiment~\cite{Bayes:2014lxz} for
$\SI{13}{\mega\eV}<m_X<\SI{80}{\mega\eV}$ with $\BR(\mueX)<\num{9e-6}$
at 90\%\,CL on average.

The characteristic signature of \mueX decays is a mono-energetic
positron whose energy is determined by the mass $m_X$ of the
undetected particle $X$. These positrons would appear as a narrow peak
on top of the smooth momentum spectrum of positrons from Standard
Model muon decays. In contrast to the \mueee search, the final state
contains only a single positron and would thus not pass event
filtering on the online GPU farm.  Therefore, a dedicated search
strategy is required, and the analysis is performed on histograms
which are filled with track fit information as part of the online
track reconstruction.  Since the full track reconstruction of all tracks
in every event frame is performed online, histograms of the total
momentum as well as the azimuthal and polar angle of the decay
electrons and positrons can be recorded.  This unprecedented dataset
of the order of $\num[]{e15}$ $\mu^{+}$ decays allows not only
for \mueX searches but also for studies of Standard Model $\mu^{+}$
decays.  As a drawback of this approach, event-by-event information is
lost and the offline reprocessing of track reconstruction is not
possible so that the optimum momentum resolution of recurling tracks
cannot be achieved.

The acceptance of the Mu3e detector determines the mass reach of the
search.  With a minimum $p_{\text{T}}(e)$ of about \SI{10}{\MeV} for
the positron to be reconstructed, $m_X$ of at most \SI{95}{\MeV} can
be studied.  It is worth noting that complementary experiments
sensitive to higher $m_X$ up to the muon mass are currently being
discussed~\cite{Koltick:2021slp}.  Low $m_X$ might be out of reach as
well, as the characteristic edge of the momentum spectrum of Michel
decays \muenunu is currently used for calibration.  Alternative
calibration methods based on Bhabha and Mott scattering are under
investigation.

\begin{figure}
    \centering
    \subfloat[][Simulated background events.] {
    \includegraphics[width=0.49\textwidth]{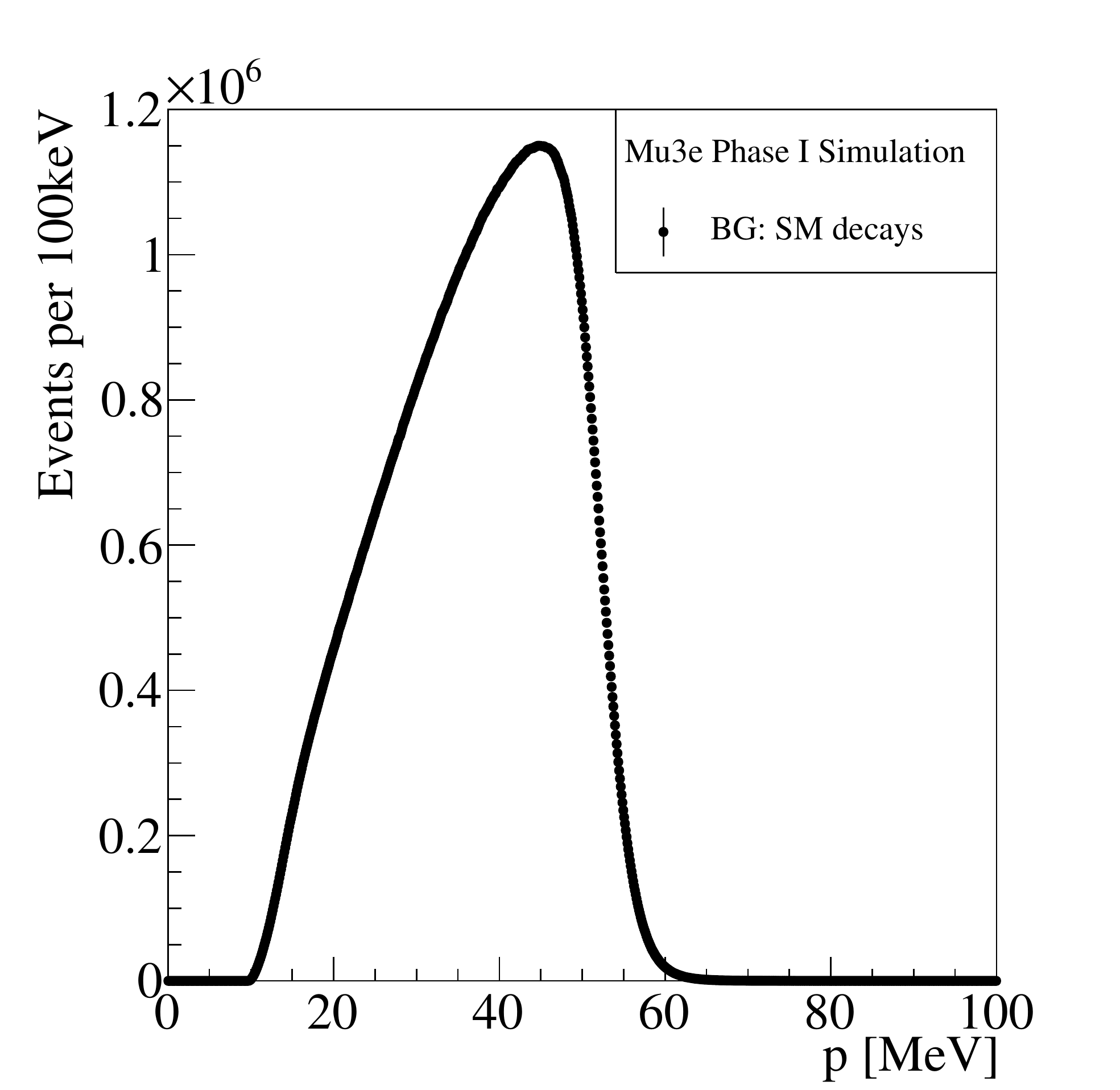}\label{fig:mueX_BG}}
    \subfloat[][Simulated \mueX signal events with $m_X=\SI{60}{\mega\eV}$.]{ 
    \includegraphics[width=0.49\textwidth]{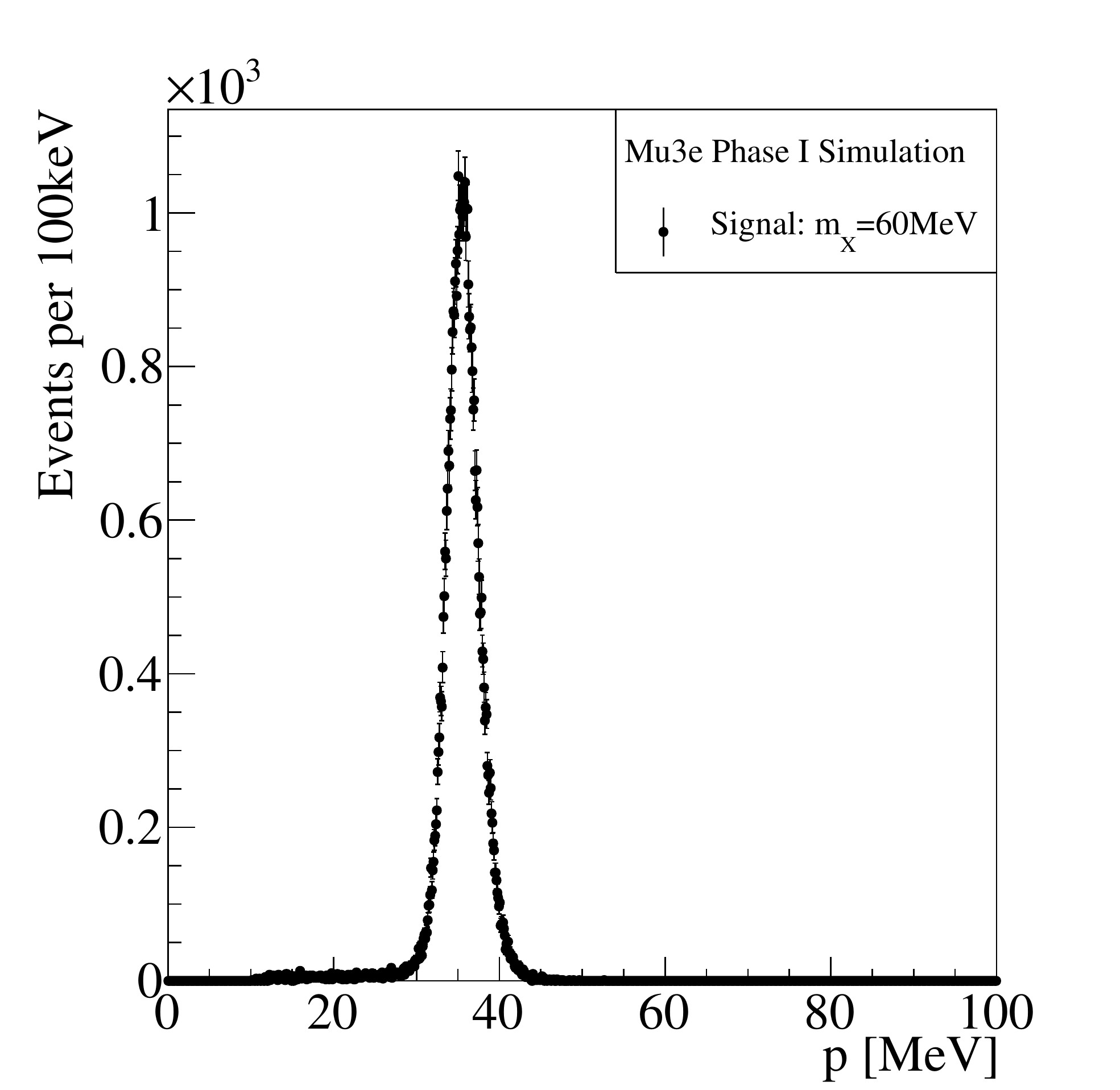}\label{fig:mueX_sig}}
    \caption{Spectra of the reconstructed positron momentum of simulated background events from Standard Model muon decays (background) and \mueX events with $m_X=\SI{60}{\mega\eV}$. The tracks are reconstructed from 4 hits in the central detector like it is done on the GPU filter farm. } 
    \label{fig:mueX_sigBG}
\end{figure}

\begin{figure}
    \centering
    \includegraphics[width=0.75\textwidth]{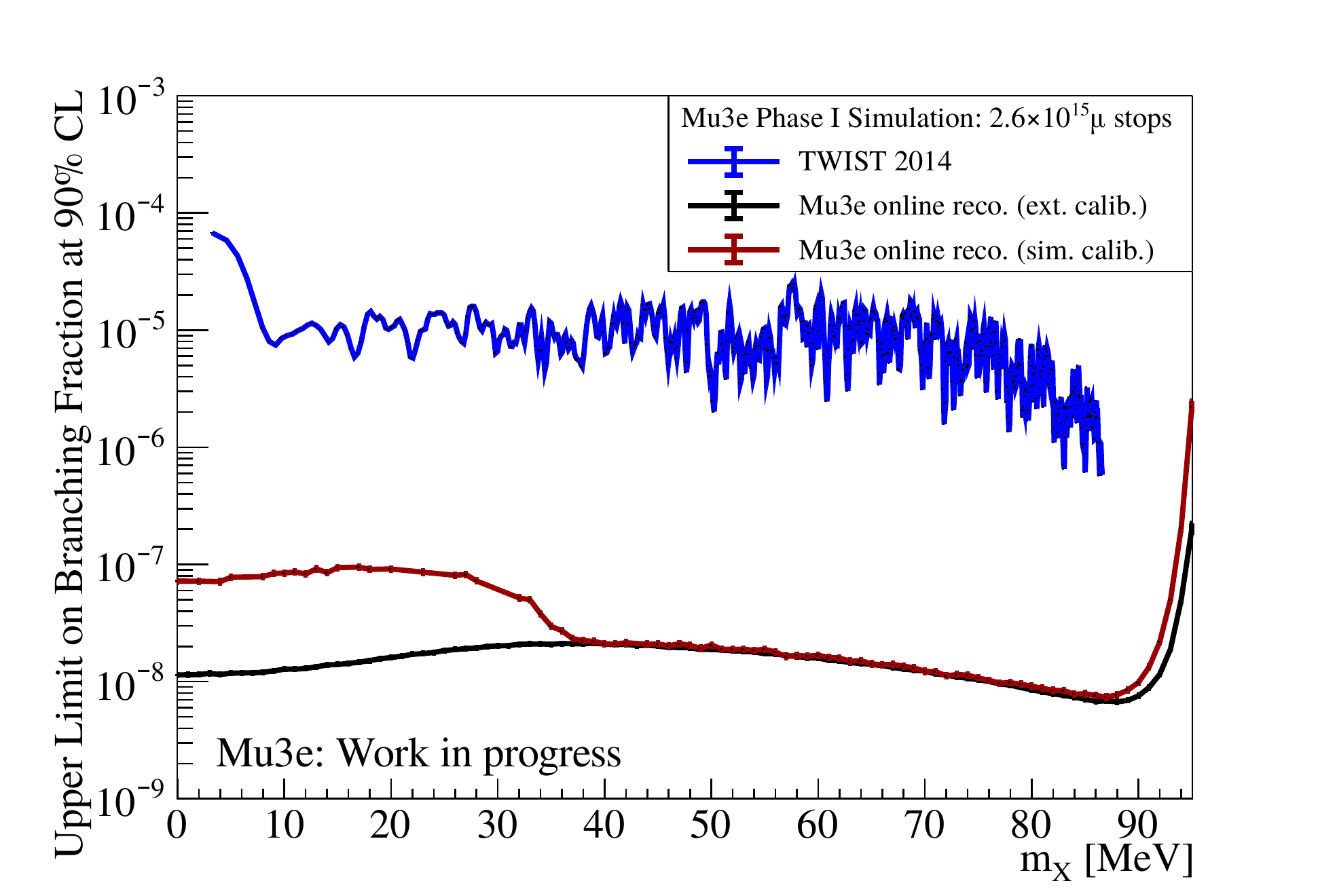}
    \caption{Expected limits on the branching fraction \BR(\mueX) at \SI{90}{\percent} CL for various masses $m_X$. 
    The momentum calibration is either obtained from the same spectrum leaving out a window around the \mueX signal (red line), or from another process such as Bhabha or Mott scattering (black line).
    The current strongest observed limits by the TWIST experiment~\cite{Bayes:2014lxz} are shown for comparison (blue line). 
    TWIST results by courtesy of R.\,Bayes.}
    \label{fig:mueX_limits}
\end{figure}

The sensitivity in Phase~1 of the Mu3e experiment is estimated in toy
Monte Carlo studies based on simulated momentum spectra.  Signal and
background events are simulated with the Geant4-based simulation of
the Mu3e experiment.  Standard Model muon decays form the background
dominated by the Michel decay \muenunu.  Additional effects like
tracks from Bhabha scattering or tracks which recurl multiple times
are also considered as background contributions.  Signal decays are
simulated as two-body decays. \\
The tracks are reconstructed from 4 hits as in the online track reconstruction. 
The reconstructed momentum spectra for background events and \mueX signal events with an exemplary mass of $m_X=\SI{60}{\mega\eV}$ are shown in Figure~\ref{fig:mueX_sigBG}. \\
In Figure~\ref{fig:mueX_limits}, the expected sensitivity of the Phase~1 Mu3e experiment is shown for two different calibration approaches.
In the first scenario, the calibration is derived from the Michel spectrum. In this approach, a momentum window around the the expected \mueX signal is left out during the calibration step.
In the second scenario, it is assumed that the calibration is derived in an alternative way for example with Bhabha or Mott scattering events. 
In the first scenario, the sensitivity is deteriorated when the left out momentum window covers the Michel edge, i.e.\,at low $m_X$. \\
In the first phase of the Mu3e experiment, \mueX decays with branching ratios in the order of \num[]{e-8} can be tested, an improvement in sensitivity by a factor of around 600 with respect to the results by the TWIST experiment~\cite{Bayes:2014lxz}. 
In addition to increased statistics in Phase~2, this search could further profit from improvements in momentum resolution. 
The feasibility of online reconstruction of long 6 or 8 hit tracks in view of the Phase~2 upgrade is currently being investigated. 

\FloatBarrier

\section{The search for $\mu^+\to e^+ + $ long-lived particles}
\label{sec:mueLLP}
The Mu3e dataset will allow searches for muon decays to light
pseudoscalar particles that are long-lived and decay within the first
silicon layer. An exploratory analysis documented in
Ref.~\cite{Heeck:2017xmg} has shown that such a final state in Mu3e
has competitive sensitivity to other experiments. This analysis,
however, has not used a detailed simulation of the Mu3e detector. In
this section, this study is repeated using the latest Mu3e detector
simulation and realistic track and vertex selection criteria.

In Ref.~\cite{Heeck:2017xmg} axion-like particles, $a$, are considered with couplings to the 
SM leptons $\ell_\alpha$ with $\alpha = e, \mu, \tau$ parameterized as:
\begin{equation}
    \mathcal{L}_a ~=~ \frac{1}{\Lambda} ~\partial_\mu a ~ \bar \ell _\alpha \gamma^\mu 
    (g^V_{\alpha\beta} + g^A_{\alpha\beta}\gamma^5) \ell_\beta ~ ,
\end{equation}
where $\Lambda$ is an effective energy scale and $g^V$, $g^A$ are the current structure matrices.
From this interaction term, the decay width of the muon decay $\mu \to e a$ and, hence, its
branching ratio, is found to be proportional to $m_\mu^3 / (16\pi \Lambda^2_{\mu e})$ 
where $\Lambda_{e\mu} = \Lambda / \sqrt{ (g^V_{e\mu})^2 + (g^A_{e\mu})^2 }$ with the
proportionality factor being a function of $m_a$ and $m_e$ under the approximation that 
$m_\mu \gg m_e$. The current structure matrices are assumed that they are such that
the axion-like particle $a$ decays only to $ee$ with a lifetime that is calculated from the
width $\Gamma(a\to ee)$ and is found to be proportional to the particle  masses
and $\Lambda_{ee} = \Lambda / g^A_{ee} $. Therefore, the free parameters for this model
are (i) the mass of the axion-like particle $m_a$, (ii) $\Lambda_{e\mu}$
that controls the branching ratio of the decay, and (iii) $\Lambda_{ee}$ that controls the
lifetime of $a$ and hence the location of its decay vertex inside the detector.

Muon decays to axion-like particles $a$ are generated using\footnote{The authors would like to thank Andrea Thamm for providing help with the {\texttt MadGraph5\_aMC@NLO} cards for this study.} {\texttt MadGraph5\_aMC@NLO}~\cite{Alwall:2014hca} 
and the resulting decay products are reconstructed using the Mu3e detector simulation.
The same selection criteria for tracks and vertices are used as in the standard $\mu\to eee$
signal search, described in Ref.~\cite{Mu3e:2020gyw}. 

\begin{figure}
    \centering
    \subfloat[]{\label{fig:mueLLP-1-a}
    \includegraphics[width=0.49\textwidth]{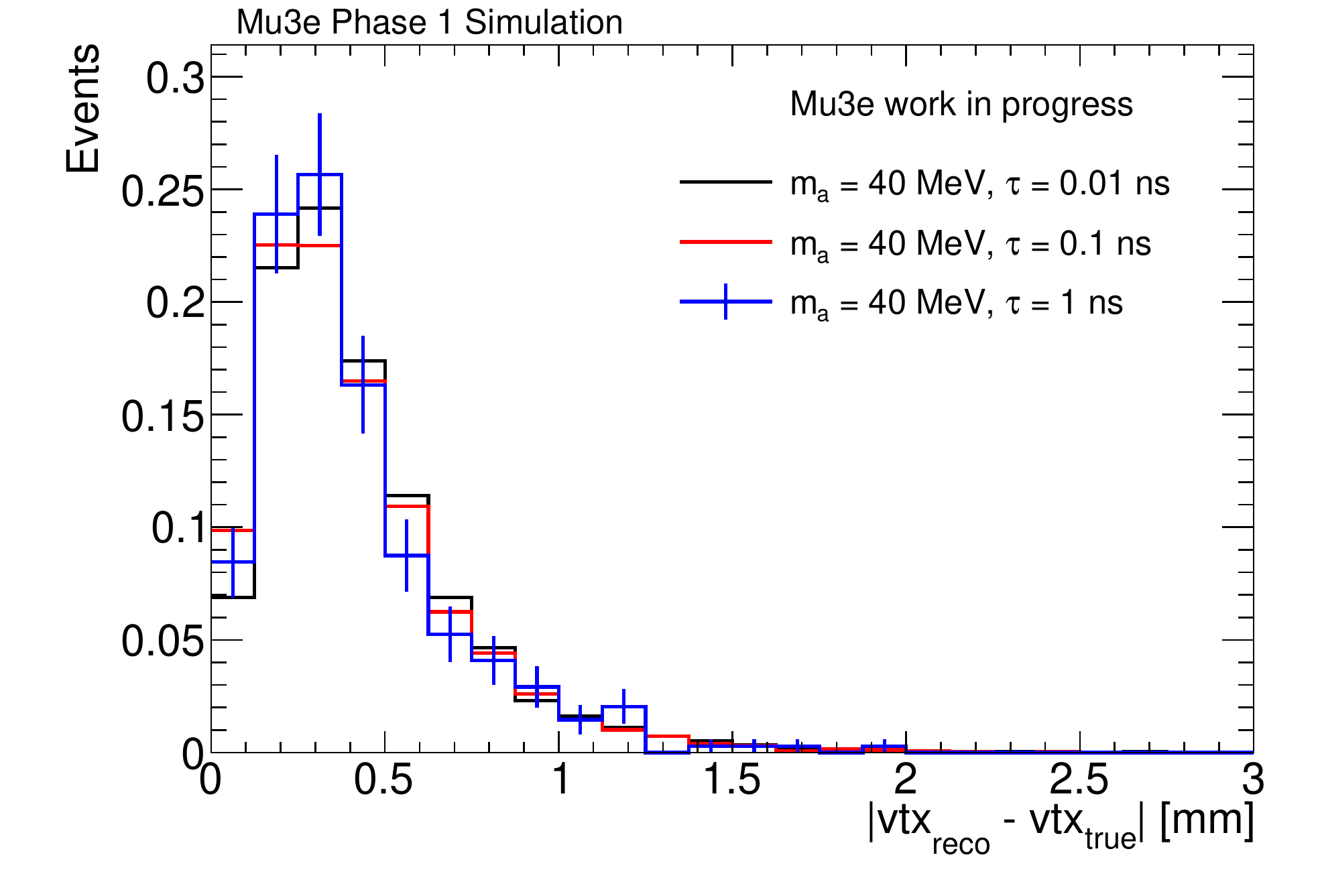}}
    \subfloat[]{\label{fig:mueLLP-1-b}
    \includegraphics[width=0.49\textwidth]{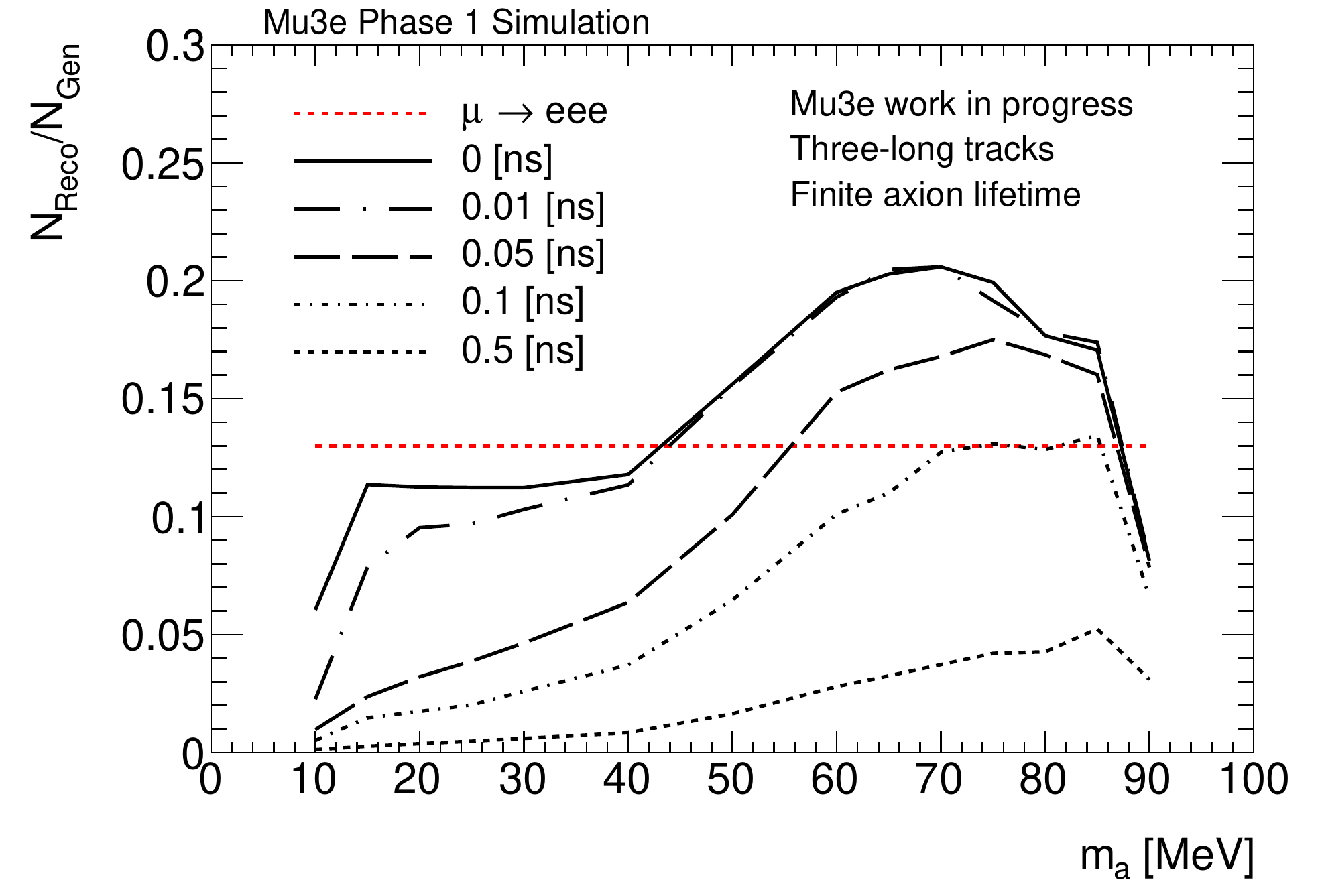}}
    \caption{The distance between the reconstructed vertex of the $e^+e^+e^-$ system using the standard Mu3e experiment vertex reconstruction algorithm and the true vertex of the axion-like particle is shown in (a). This is for an axion-like particle of mass 40~MeV and lifetimes of 0.01~ns, 0.1~ns and 1~ns. The efficiency of the standard Mu3e experiment signal selection for axion-like particle masses in the range 10--90 MeV and for a range of lifetimes is shown in (b). } 
    \label{fig:mueLLP-1}
\end{figure}

In this particular decay, the  muon is assumed to stop on the target and to decay  back-to-back
to an electron and an axion-like particle $a$. The $a$ particle will have a given boost and will
decay depending on the assumed lifetime to two electrons. Due to the topology of the decay,
the electron from the muon decay has a track that, if extrapolated, will meet the $a$ particle's
decay vertex. Therefore, for as long as there is no requirement for the reconstructed vertices
to be close to the target surface, the Mu3e vertex reconstruction should be very efficient.
The distance of the reconstructed vertex of the $eee$ system from the actual $a$ particle
decay vertex for long tracks selected for this analysis is shown in Fig.~\ref{fig:mueLLP-1-a}. This distribution
is compatible with the vertex resolution performance for $\mu\to eee$ decays. 
The efficiency on signal of various $m_a$ and lifetime values, as well as the efficiency
for $\mu\to eee$ decays for comparison, is shown in Fig.~\ref{fig:mueLLP-1-b}. No requirement on the distance
of the vertices from the target is applied for these results.

\begin{figure}
    \centering
    \label{fig:mueLLP-2-a}
    \includegraphics[width=0.49\textwidth]{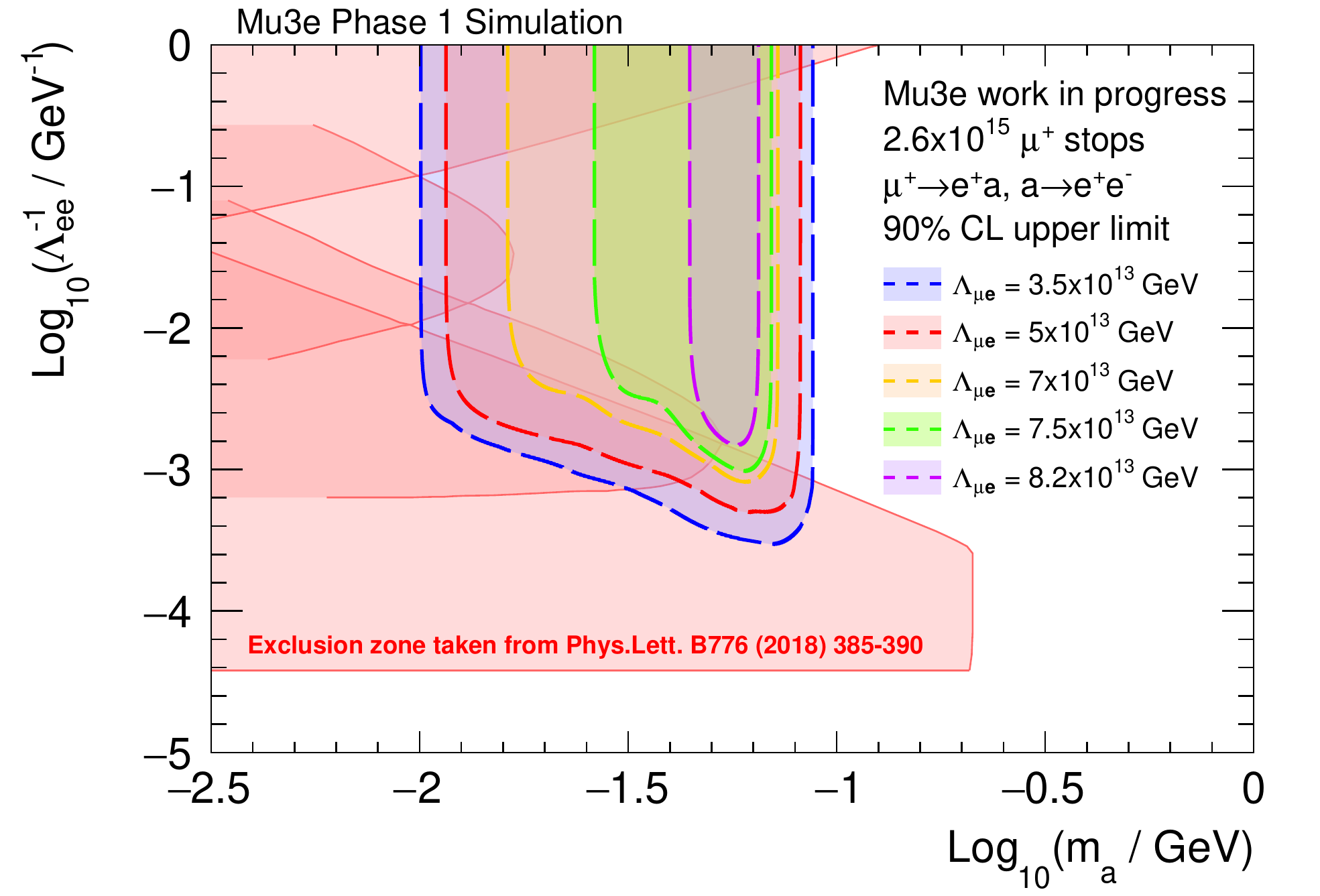}
    \caption{Upper limits in the two dimensional parameter space of the simple
    axion-like particle model discussed in the text. 
    } 
    \label{fig:mueLLP-2}
\end{figure}

The efficiencies shown in Fig.~\ref{fig:mueLLP-1-b} are interpreted in
the $m_a$--$\Lambda_{ee}$ plane for different values of
$\Lambda_{e\mu}$ assuming the Mu3e Phase~1 expected muon decays on
target as a 90\% confidence level upper limit assuming no signal is
found in a zero background experiment.  This is shown in
Fig.~\ref{fig:mueLLP-2-a}. On the same figure, exclusions from beam
dump experiments and electron $g-2$ reproduced from
Ref.~\cite{Heeck:2017xmg} are also shown.  The whole available region
of the $m_a$--$\Lambda_{ee}$ parameter space between the existing
constraints and the muon mass kinematic cutoff is excluded for
$\Lambda_{e\mu} < 5\times 10^{13}$~GeV.  To reduce combinatoric
backgrounds, a requirement on the distance between the vertex and
target my be required. The impact of a ealistic value of 3~mm (suggested in Ref.~\cite{Perrevoort:2018okj}) has been studied and the outcome is that
such a requirement has a small effect on the sensitivity and does not
impair the sensitivity of the analysis to the axion parameter space.

\FloatBarrier

\section{The search for \ee-resonances in \mueeenunu}
\label{sec:muDP}
In addition to \mueee searches and searches for long-lived particles, the dataset that will be recorded by the Mu3e experiment also allows for searches for resonances in \mueeenunu. 
One example that has been further studied are dark photon \DP decays into \ee pairs. \\
The dark photon is the messenger of a potential vector portal to the dark sector.
It interacts with Standard Model particles via kinetic mixing with the photon and $Z$ boson, i.e.\ via coupling to the electro-magnetic current. 
If the dark photon is light enough, it can be radiated in muon decays: \mueAnunu. 

In the following, the Lagrangian from~\cite{Echenard:2014lma} is used with the parameter $\epsilon$ determining the strength of the kinetic mixing. 
In this study, the sensitivity of the Mu3e experiment to promptly decaying dark photons \Aee emitted in muon decays is estimated with a simplified detector simulation and without considering background from Bhabha scattering events. 
In the study presented in the following, the full Mu3e detector simulation and event reconstruction are used. \\
Signal events are generated with {\texttt MadGraph5\_aMC@NLO} 2.4.3~\cite{Alwall:2014hca} up to dark photon masses of $\mA=\SI{80}{\MeV}$. 
At larger masses, the parameter space in reach of Mu3e is already excluded by existing experimental limits. 
The simulated dark photons decay promptly to \ee so that the same track and vertex reconstruction as for the \mueee search can be employed. \\
The background is dominated by the rare muon decay \mueeenunu.
As additional background sources, accidental combinations of Bhabha scattering events with Michel decays are considered which contribute on average by a factor of 800 less than the rare muon decay.
Further types of accidental background are even less likely and therefore not considered in this study. 

\begin{figure}
    \centering
    \subfloat[][Simulated background events. Both combinations of \ee are considered. ] {
    \includegraphics[width=0.475\textwidth]{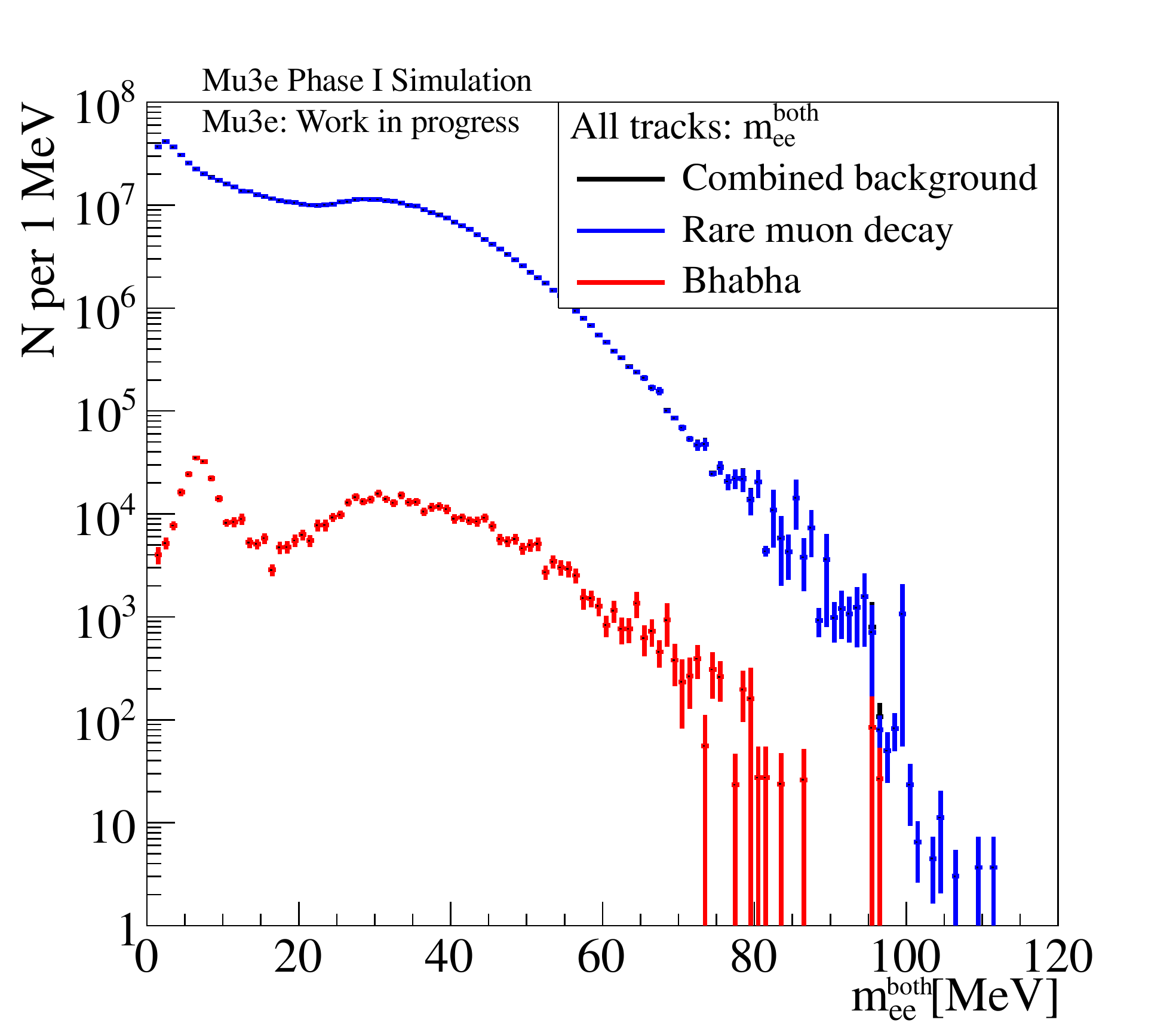}\label{fig:muDP_BG}}\quad
    \subfloat[][Simulated \DP signal events. Both combinations of \ee are considered. ]{ 
    \includegraphics[width=0.475\textwidth]{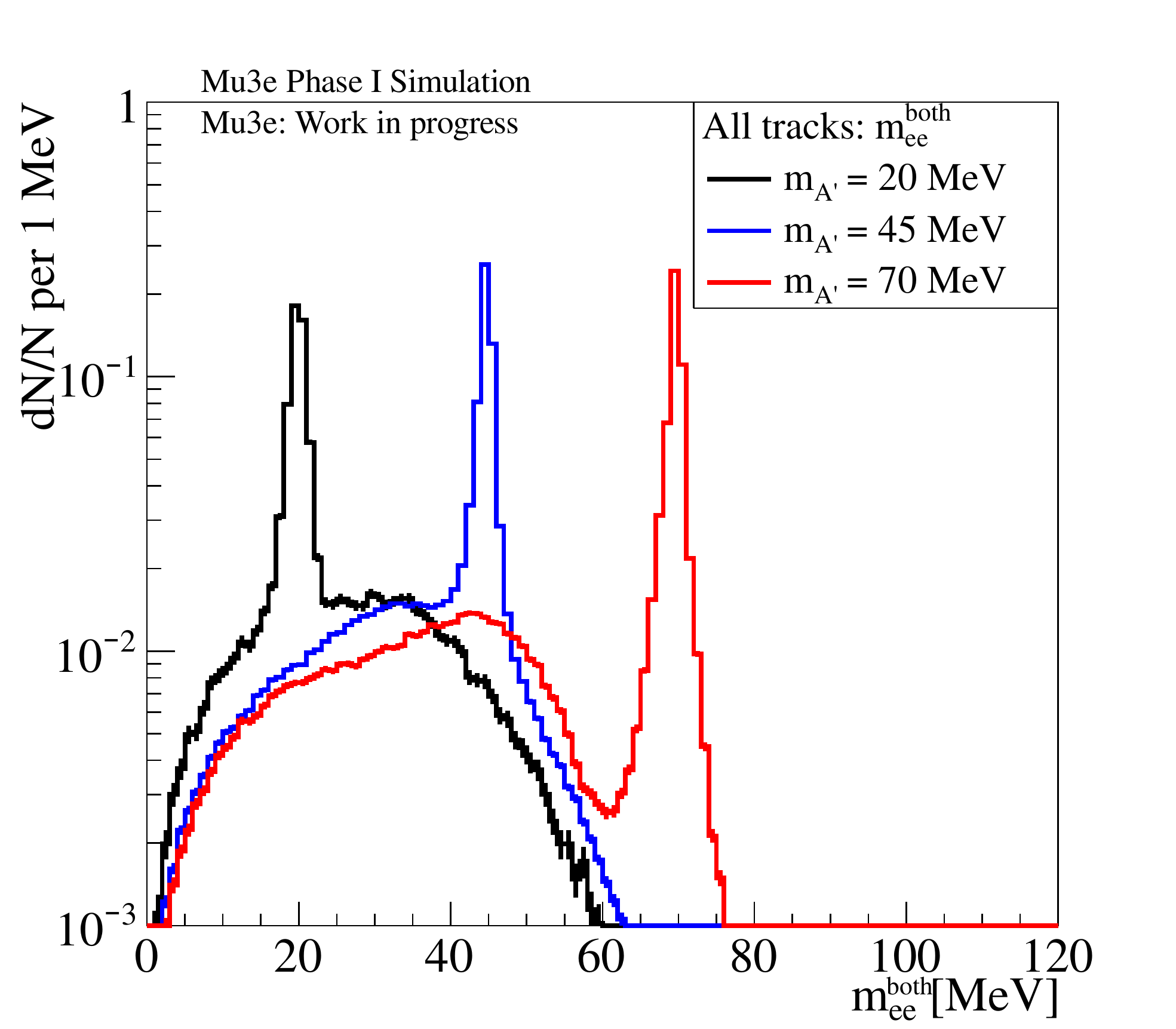}\label{fig:muDP_sigAll}}\\
    \subfloat[][Simulated \DP events. The \ee-combination with the lower invariant mass is shown. ] {
    \includegraphics[width=0.475\textwidth]{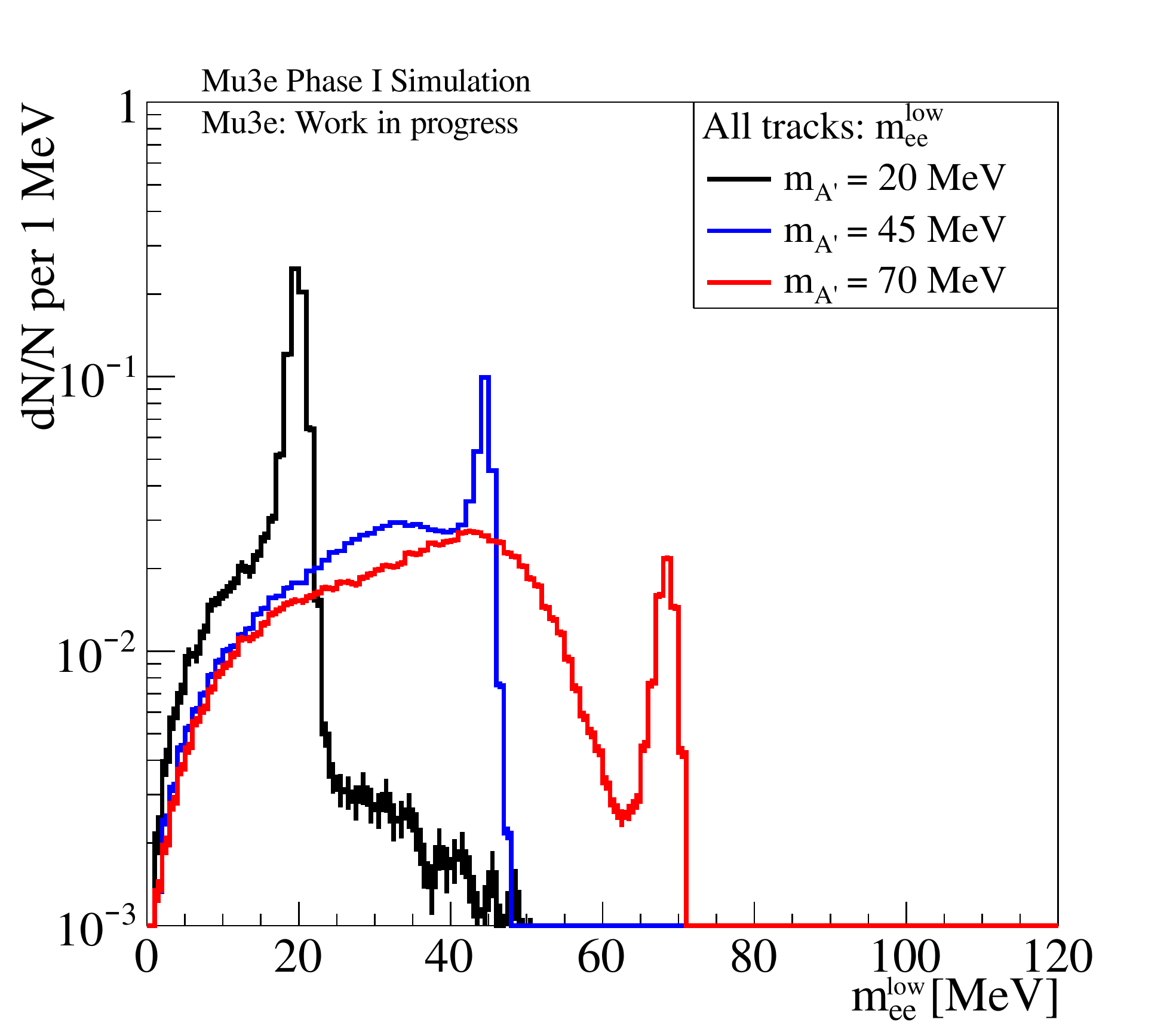}\label{fig:muDP_sigLow}}\quad
    \subfloat[][Simulated \DP signal events. The \ee-combination with the higher invariant mass is shown. ]{ 
    \includegraphics[width=0.475\textwidth]{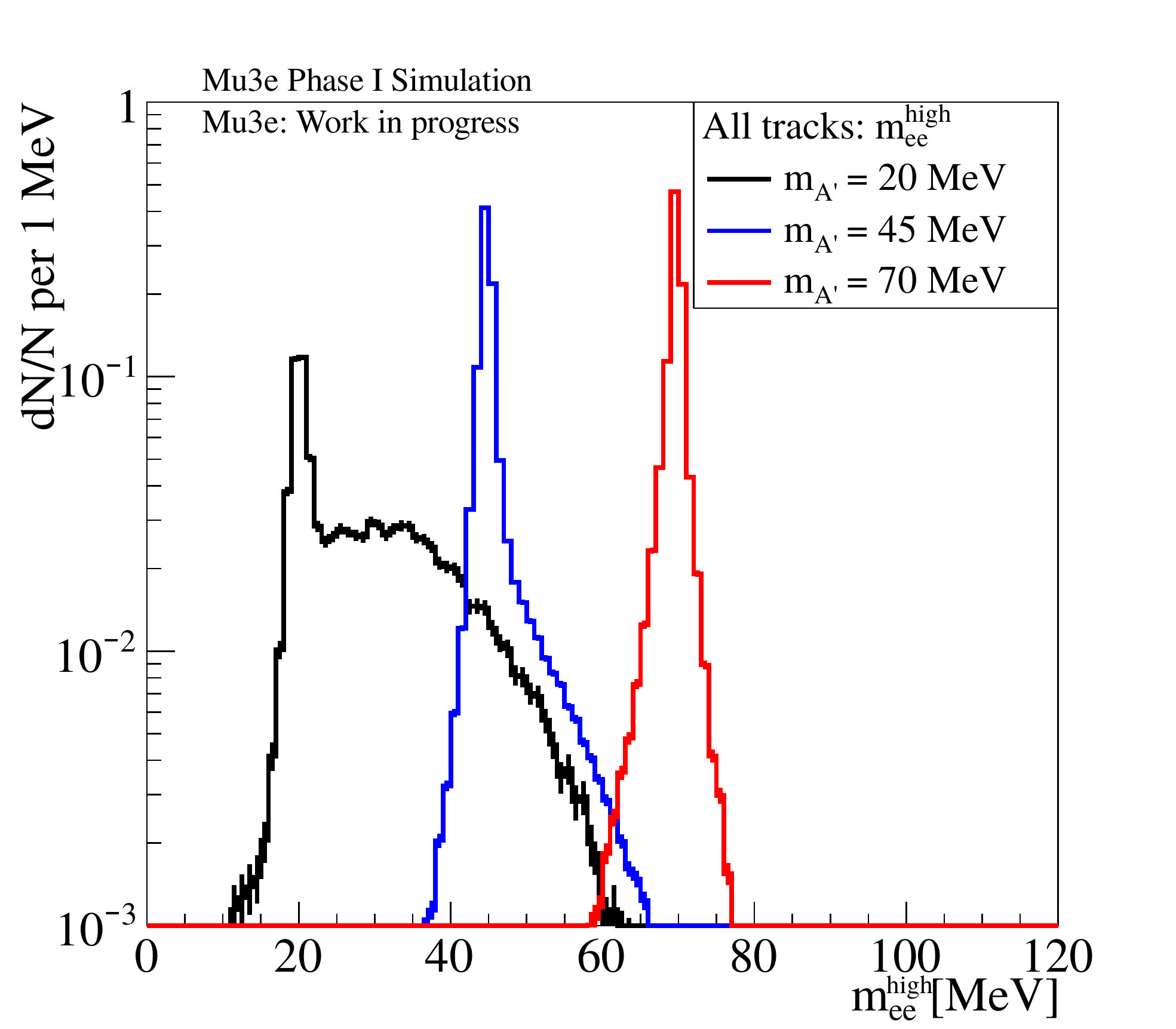}\label{fig:muDP_sigHigh}}
    \caption{Spectra of the reconstructed invariant mass $m_{ee}$ of simulated background and dark photon signal events with $\mA=\SI{20}{\MeV}, \SI{45}{\MeV}$ and $\SI{70}{\MeV}$. }
    \label{fig:muDP_sigBG}
\end{figure}

Spectra of the reconstructed invariant \ee-mass $m_{ee}$ for background and signal are shown in Figure~\ref{fig:muDP_sigBG}. 
The signal distribution features a narrow peak on top of a broader distribution which stems from \ee-combination of the positron from the muon decay and the electron from the \DP decay. 
This contribution can be reduced by selecting the \ee-combination with the lower $m_{ee}$ for lighter \DP, and the combination with the higher $m_{ee}$ for heavier \DP. 
The optimum transition point lies around \SI{45}{\MeV}. 

\begin{figure}
    \centering
    \subfloat[Expected limits on the branching fraction at 90\% CL. ]{\includegraphics[height=0.225\textheight]{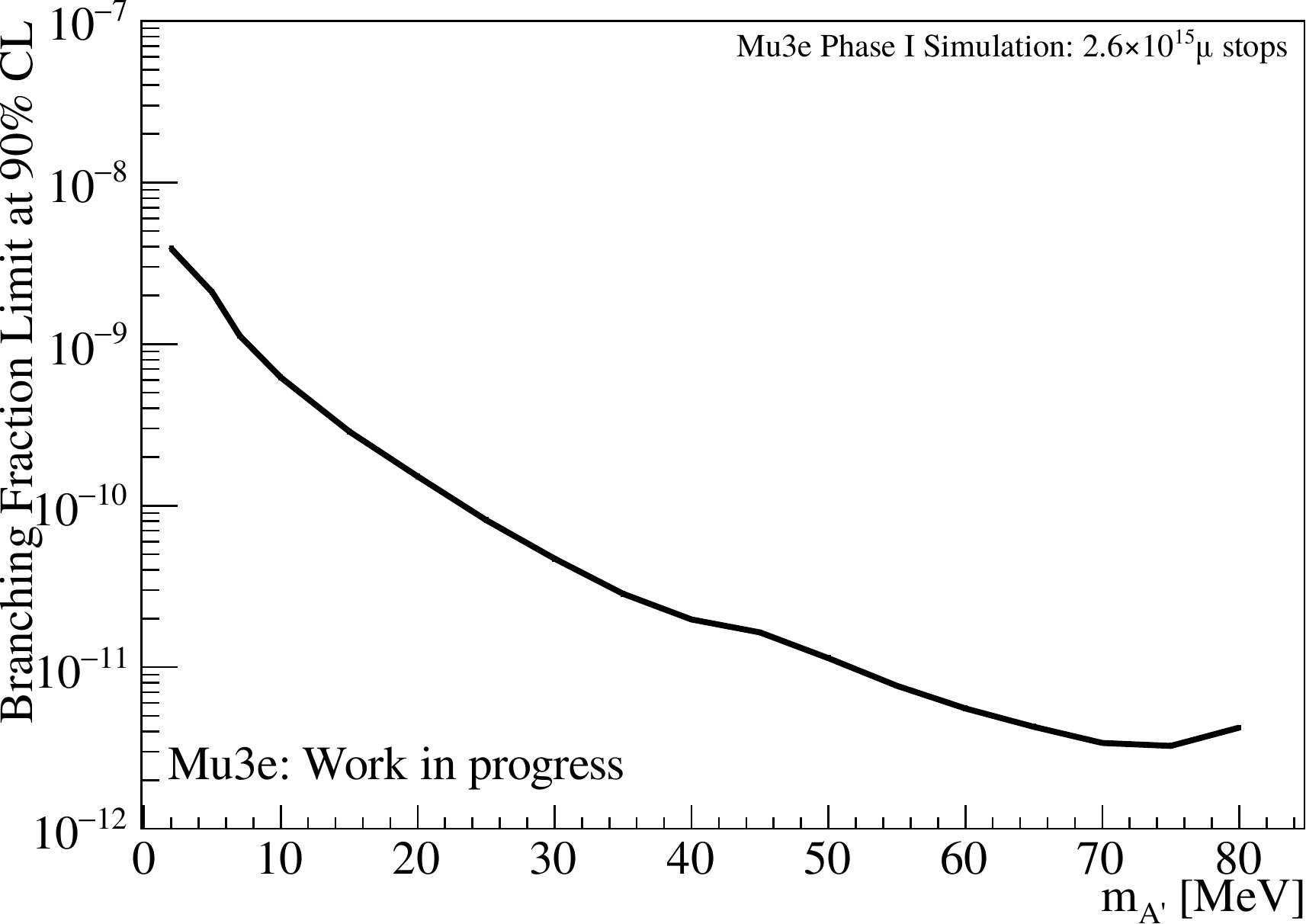}\label{fig:muDP_limitsBR}}\quad
    \subfloat[Expected limits on the kinetic mixing parameter $\epsilon$ at 90\%CL. Adapted from~\cite{Ablikim:2017aab}. ]{\includegraphics[height=0.225\textheight]{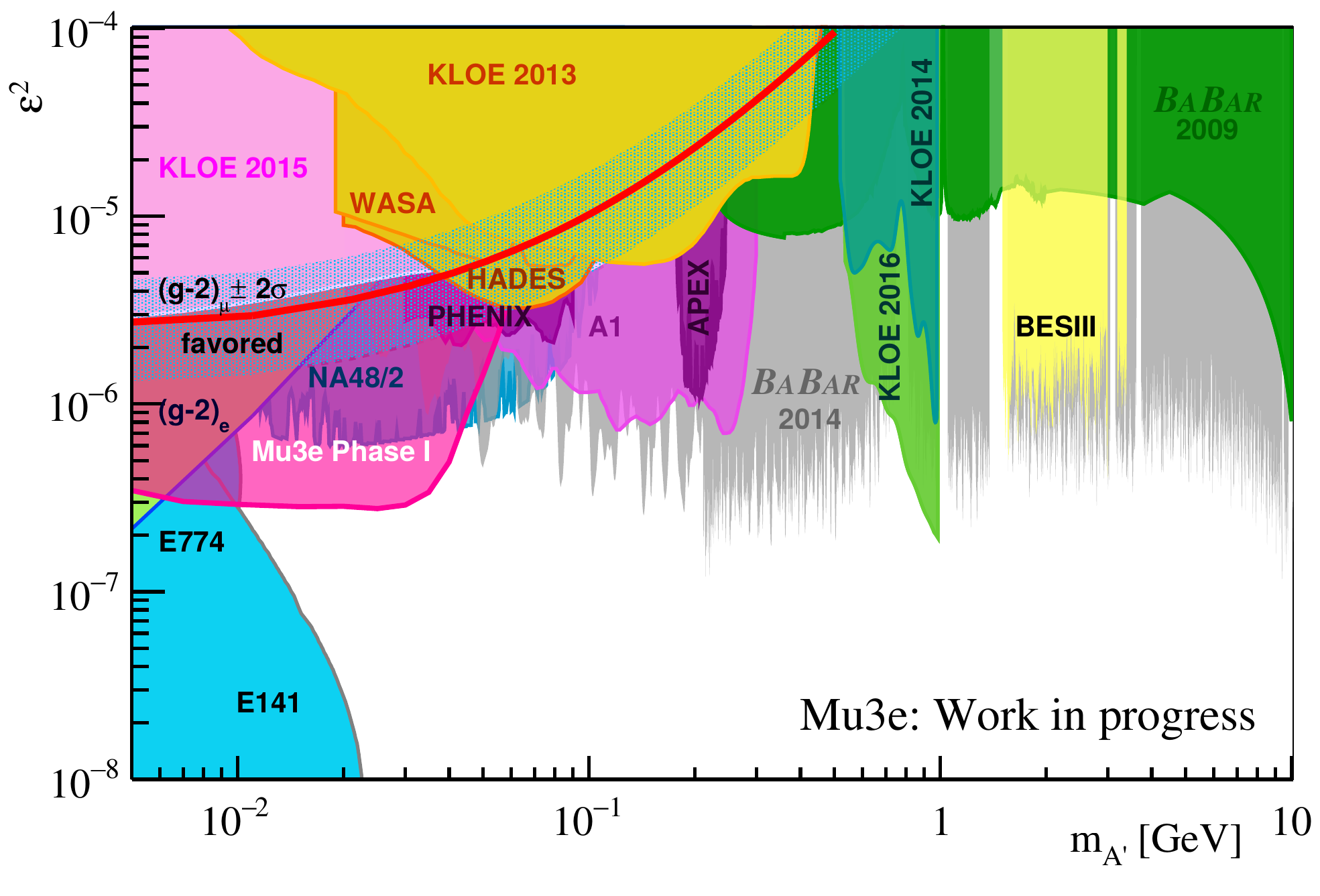}\label{fig:muDP_limitsComp}}
    \caption{Expected sensitivity to prompt dark photon decays in \mueeenunu in the first phase of the Mu3e experiment. }
    \label{fig:muDP_limits}
\end{figure}

The sensitivity to prompt dark photon decays in muon decays in the Phase~1 Mu3e experiment is estimated with toy Monte Carlo studies. 
At low \mA, branching fractions of \num{5e-9} at 90\% CL can be investigated, while at higher \mA, branching fraction of \num{3e-12} at 90\% CL can be reached (see Figure~\ref{fig:muDP_limitsBR}). 
The expected limits on the branching fraction are translated into limits on the kinetic mixing parameter $\epsilon$. 
The reach in the $\epsilon-\mA$ parameter of the Phase~1 Mu3e experiment is shown alongside observed limits is shown in Figure~\ref{fig:muDP_limitsComp}.

\FloatBarrier

\section{\label{sec::conclusion}Conclusion}
The Mu3e experiment will search for the lepton-number-violating decay
\mueee. Using a two-phase approach and innovative design, it will
extend the sensitivity to the branching ratio for this process to $10^{-16}$, four
orders of magnitude beyond current limits. In the context of the global
programme covering \mueee, \meg and \muconv, Mu3e provides unique and
complementary sensitivity to flavour-violating new physics.

The Mu3e experimental design also allows the search for a range of
other processes, including dark photons, axion-like particles and
long-lived particles. In each case, Mu3e can extend existing
sensitivities in the accessible range of parameter space for such
models.

\section*{\label{sec::acknowledgments}Acknowledgements}
The UK institutes thank the Science and Technology Facilities Council
for funding their work through the Large Projects scheme, under grant
numbers: ST/P00282X/1, ST/P002765/1, ST/P002730/1, ST/P002870/1.
A. Perrevoort's work is funded by the Federal Ministry of Education
and Research (BMBF) and the Baden-W\"urttemberg Ministry of Science as
part of the Excellence Strategy of the German Federal and State
Governments.  A. Perrevoort further acknowledges the support by the
German Research Foundation (DFG) funded Research Training Group
“Particle Physics beyond the Standard Model” (GK 1994) on previous
works on this subject.  \bibliographystyle{JHEP} \bibliography{main}

\end{document}